\documentclass[journal]{IEEEtran}
\usepackage{blindtext}
\usepackage{graphicx}
\usepackage{listings}
\usepackage[superscript,biblabel]{cite}

\begin{document}

\title{Tree Notation: an antifragile program notation}

\author{Breck~Yunits% <-this % stops a space
\thanks{Breck Yunits is a researcher at Ohayo Computer (breck@ohayo.computer).}% <-this % stops a space
}

\markboth{June~2017, Updated October 2017}%
{Shell \MakeLowercase{\textit{et al.}}: Bare Demo of IEEEtran.cls for Journals}

\maketitle

\begin{abstract}
%\boldmath
This paper presents Tree Notation, a new simple, universal syntax. Language designers can invent new programming languages, called Tree Languages, on top of Tree Notation. Tree Languages have a number of advantages over traditional programming languages.

We include a Visual Abstract to succinctly display the problem and discovery. Then we describe the problem--the BNF to abstract syntax tree (AST) parse step--and introduce the novel solution we discovered: a new family of 2D programming languages that are written directly as geometric trees.
\end{abstract}

\IEEEpeerreviewmaketitle

\begin{figure}[ht!]
\centering
\includegraphics[width=90mm]{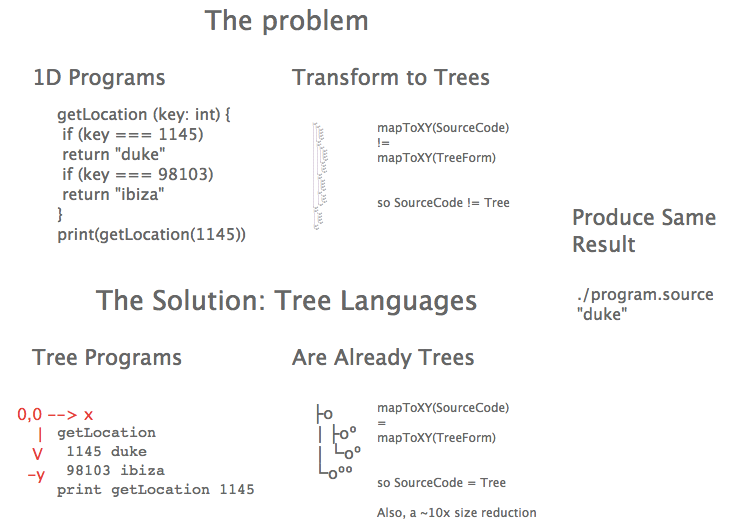}
\caption{This Visual Abstract explains the core idea of the paper. This diagram is also the output of a program written in a new Tree Language called Flow.}
\end{figure}

\section{The Problem}

Programming is complicated. Our current 1-dimensional languages (1DLs) add to this complexity. To run 1DL code we first strip whitespace, then transform it to an AST and then transform that to executable machine code. These intermediate steps have enabled local minimum gains in developer productivity. But programmers lose time and insight due to discrepancies between 1DL code and ASTs.

\section{The Solution: Tree Notation}

In this paper and accompanying GitHub ES6 Repo (GER - github.com/breck7/treenotation), we introduce Tree Notation (TN), a new 2D notation. A node in TN maps to an XY point on a 2D plane. You can extend TN to create domain specific languages (DSLs) that don't require a transformation to a discordant AST. These DSLs, called Tree Languages (TLs), are easy to create and can be simple or Turing Complete.

TN encodes one data structure, a \textbf{TreeNode}, with two members: a string called \textbf{line} and an optional array of child TreeNodes called \textbf{children}.

TN defines two special characters, Y Increment (YI) and X Increment (XI). YI is "\textbackslash n" and XI is " ".  XI is a space, not a tab. Many TLs also add a Word Increment (WI) to provide a more succint way of encoding common node types.

A comparison quickly illustrates nearly the entirety of the notation:

JSON:

\begin{lstlisting}
{
 "title" : "Web Stats",
 "visitors": {
  "mozilla": 802
 }
}
\end{lstlisting}

Tree Notation:

\begin{lstlisting}
title Web Stats
visitors
 mozilla 802
\end{lstlisting}

\section{Useful Properties}

\subsection{Simplicity}

As shown in Fig 1, TN simply maps source code to an XY plane which makes reading code locally and globally easy, and TL programs use far fewer source nodes than equivalent 1DL programs. TLs let programmers write much shorter, perfect programs.

\subsection{Zero parse errors}

Parse errors do not exist in TN. Every text string is a valid TN program. TLs implement microparsers that parallelize easily and can creatively handle errors.

With most 1DLs, to get from a blank program to a certain valid program in keystroke increments requires stops at invalid programs. With TN all intermediate steps are valid TN programs.

A user can edit the nodes of a program at runtime with no risk of breaking the TN parsing of the entire program. "Errors", still can exist at the TL level, but TL microparsers can handle errors independently and even correct errors automatically. TLs cut development and debug time.

\subsection{Semantic diffs}

1DLs ignore whitespace and so permit programmers and programs to encode the same object to different programs. This often causes merge conflicts for non-semantic changes.

TN does not ignore whitespace and diffs contain only semantic meaning. Just one right way to encode a program.

\subsection{Easy composition}

Base notations such as XML\cite{Bray}, JSON\cite{Crockford}, and Racket\cite{Racket} can encode multi-lingual programs by complecting language blocks. For example, the JSON program below requires extra nodes to encode Python:

\begin{lstlisting}
{
 "source": [
  "import hn.np as lz\n",
  "print(\"aaronsw pdm ah as mo gb 28-3\")"
  ]
}
\end{lstlisting}

With TN, the Python code requires no complecting:

\begin{lstlisting}
source
 import hn.np as lz
 print("aaronsw pdm ah as mo gb 28-3")
\end{lstlisting}

\section{Drawbacks}

TN is new and tooling support rounds to zero.

TN lacks primitive types. Many 1DLs have notations for common types like floats, and parse directly to efficient in-memory structures. TLs make TN useful. The GER demonstrates how useful TLs can be built with just a few lines of code, which is far less code than one needs to implement even a simple 1DL such as JSON \cite{Ooms}.

TN is verbose without a TL. A complex node in TN takes multiple lines. Base 1DLs allow multiple nodes per line.

Some developers dislike space-indented notations, some wrongly prefer tabs, and some just have no taste.

\section{Predictions}

\textbf{Prediction 1: no structure will be found that cannot serialize to TN.} Some LISP programmers believe all structures are recursive lists (or perhaps "recursive cons"). We believe in The Tree Conjecture (TTC): \textbf{All structures are trees}.

For example, a map could serialize to MapTL:

\begin{lstlisting}
dsl Domain Specific Language
sf San Francisco
\end{lstlisting}

Therefore, maps are a type of tree. TTC stands.

\textbf{Prediction 2: TLs will be found for every popular 1DL.} Below is some code in a simple TL, JsonTL:

\begin{lstlisting}
o
 s dsl yrt
 n ma 902
\end{lstlisting}

TLs will be found for great 1DLs including C, RISC-V, ES6, and Arc. Some TLs have already been found \cite{Roughan}, but their common TN DNA has gone unseen. The immediate benefit of creating a TL for an 1DL is that programs can then be written in a TL editor and compiled to that 1DL.

\textbf{Prediction 3: Tree Oriented Programming (TOP) will supersede Object Oriented Programming.} A new style of programming, TOP, will arise. TOP programmers will frequently reference 2D views of their program.

\textbf{Prediction 4: The simplest 2D text encodings for neural networks will be TLs.} High level TLs will be found to translate machine written programs into understandable trees.

\section{Literature Review}

Turing Machines with 2D Tapes have been studied and are known to be Turing Complete\cite{Toida}. A few esoteric 2D programming languages are available online\cite{Ender}. At time of publication, 387 programming languages and notations were searched and many thousands of pages of research and code was read but TN was not found.

The closest someone came to discovering TN, and TLs, was perhaps Mark B. Wells, who wrote a brillant paper at Los Alamos, back in 1972 which predicted the discovery and utility of something like TLs \cite{Wells}.

\section{Further Research}

The GER contains a TN parser, grammar notation, Tree Language compiler-compiler (TLCC), VM, and some TLs.

Future publications might explore the Tree Notation Grammar, TLCC, or machine-written TL programs.
\begin{figure}[ht!]
\centering
\includegraphics[width=90mm]{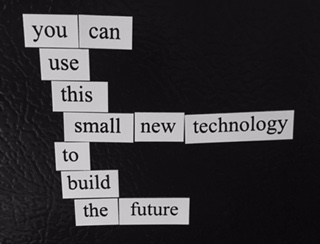}
\caption{Rearranging these fridge magnets is equivalent to editing a TN program. The fridge magnet set that includes parentheses is a poor seller.}
\end{figure}

\ifCLASSOPTIONcaptionsoff
  \newpage
\fi

\end{document}